\documentclass{ws-procs9x6}            
\begin{document}
\title{New directions with transfer reactions}

\author{Kate L. Jones$^*$ for the VANDLE, GODDESS, HAGRiD, and CLARION-BareBall Collaborations}

\address{Department of Physics and Astronomy, University of Tennessee,\\
Knoxville, TN 37996, USA\\
$^*$E-mail: kgrzywac@utk.edu\\
http://www.phys.utk.edu/faculty/faculty-jones.html}

\begin{abstract}
Over the last two decades transfer reactions have seen a resurgence following developments in methods to use them with exotic beams.  An important step in this evolution was the ability to perform the (d,p) reaction on fission fragment beams using the inverse kinematics technique, built on the experience with light beams.  There has been renewed interest in using ($^9$Be, $^8$Be) and ($^{13}$C, $^{12}$C) reactions to selectively populate single-particle like states that can be studied via their subsequent $\gamma$ decay.  These reactions have been successfully utilized in the $^{132}$Sn region.  Additionally, our collaboration has recently performed experiments with GODDESS, a combination of the full ORRUBA detector and Gammasphere arrays.  Another new direction is measuring neutrons from (d,n) reactions, performed in inverse kinematics, with the VANDLE array of plastic scintillators.  Presented below is an overview of these new techniques and some of the early data from recent experiments.
\end{abstract}


\bodymatter

\section{Introduction}
As radioactive ion beams became available with energies around or above the Coulomb barrier, techniques were developed to use them with transfer reactions in inverse kinematics with light targets, typically of protons or deuterons.  This opened up opportunities for detailed studies of exotic nuclei that were not previously available.  This approach was very successful, with many groups studying exotic nuclei from fragmentation, reaccelerated Isotope Separator OnLine (ISOL), and in-flight facilities.  Now, in the years leading up to the Facility for Rare Ion Beams (FRIB) online operations, it is a good time to develop new tools and techniques to use at FRIB.  This proceeding briefly describes some of the developments made by our collaborations towards tools and techniques that can be used at future facilities.

\section{Informing (n,$\gamma$) reaction calculations for astrophysics}

Transfer reactions are a powerful tool for understanding the structure of the nucleus, through selectively populating certain states of interest, aiding in making spin-parity ($J^\pi$) assignments, and the extraction of spectroscopic factors, and thereby testing models of the nucleus.  This information can also be used in nuclear astrophysics to inform nucleosynthesis simulations.

One application of the information derived from transfer reactions is in providing level information required for calculating direct neutron capture rates for element production.  The rapid neutron capture process, commonly termed the r-process, occurs in a neutron-rich environment and creates approximately half of the stable elements heavier than iron \cite{Bur57}.  The site of the r-process is still disputed, with the neutrino-driven wind of a core collapse supernova and neutron star mergers as two leading contenders.  Having reliable nuclear data is critical to understanding the r-process and to defining the astrophysical site.  The process proceeds through many neutron captures on seed nuclei in the nickel-iron region of the chart of nuclides, until the neutron-capture rate is in balance with that for photo-disintegration.The process can continue through $\beta$ decay, which produces a higher-Z isobar. During the main r-process, a narrow band of isotopes running diagonally through medium-mass and heavy neutron-rich nuclei will be populated owing to an equilibrium between (n,$\gamma$) and ($\gamma$,n) reactions and the short half lives of these nuclei.

At late times in the r-process, which lasts of the order of 1 second, the (n,$\gamma$) ($\gamma$,n) equilibrium is broken as the number of free neutrons that are available to induce neutron capture decreases.  During this freezeout period, individual neutron-capture rates become important to the final abundances of elements \cite{Sur01}.  A sensitivity study \cite{beu09} showed that the global r-process pattern is sensitive to the $^{130}$Sn(n,$\gamma$) rate, which not only moves material from $^{130}$Sn to $^{131}$Sn, but also shifts the production of elements for rare earth and heavier elements.  This second effect is due to the large absorption of neutrons in the case that the $^{130}$Sn(n,$\gamma$) rate is high.

Our collaboration has made a series of studies at the Holifield Radioactive Ion Beam Facility (HRIBF) \cite{Bee11} using the (d,p) reaction on beams of tin isotopes from stable $^{124}$Sn through to doubly-magic $^{132}$Sn \cite{Jon04, Man17, Koz12, Jon11}.  The unstable isotopes were produced as fragments from the proton-induced fission of uranium.  Calculations of the direct semidirect (DSD) (n,$\gamma$) rate, that is the rate including both the direct capture and the effect of the giant dipole resonance, have been made for the neutron-rich isotopes $^{124,126,128,130,132}$Sn \cite{Man17}.  These calculations relied on structure information, including spectroscopic factors, extracted from the (d,p) reactions.

In the particular case of $^{130}$Sn, the DSD (n,$\gamma$) rate was very poorly known prior to these series of measurements, as the energies of the $3p_{1/2}$ and $3p_{3/2}$ states, which are the dominant contributors to the DSD rate, were unknown, as well as the spectroscopic factors.  The HRIBF measurements reduced the uncertainty in this rate by orders of magnitude \cite{Koz12}.

\section{Measuring $\gamma$ rays following transfer reactions}
	\subsection{Incorporating large $\gamma$ arrays}
	
\begin{figure} [h]
\hspace*{-0.2in}\includegraphics[width=5.1in] {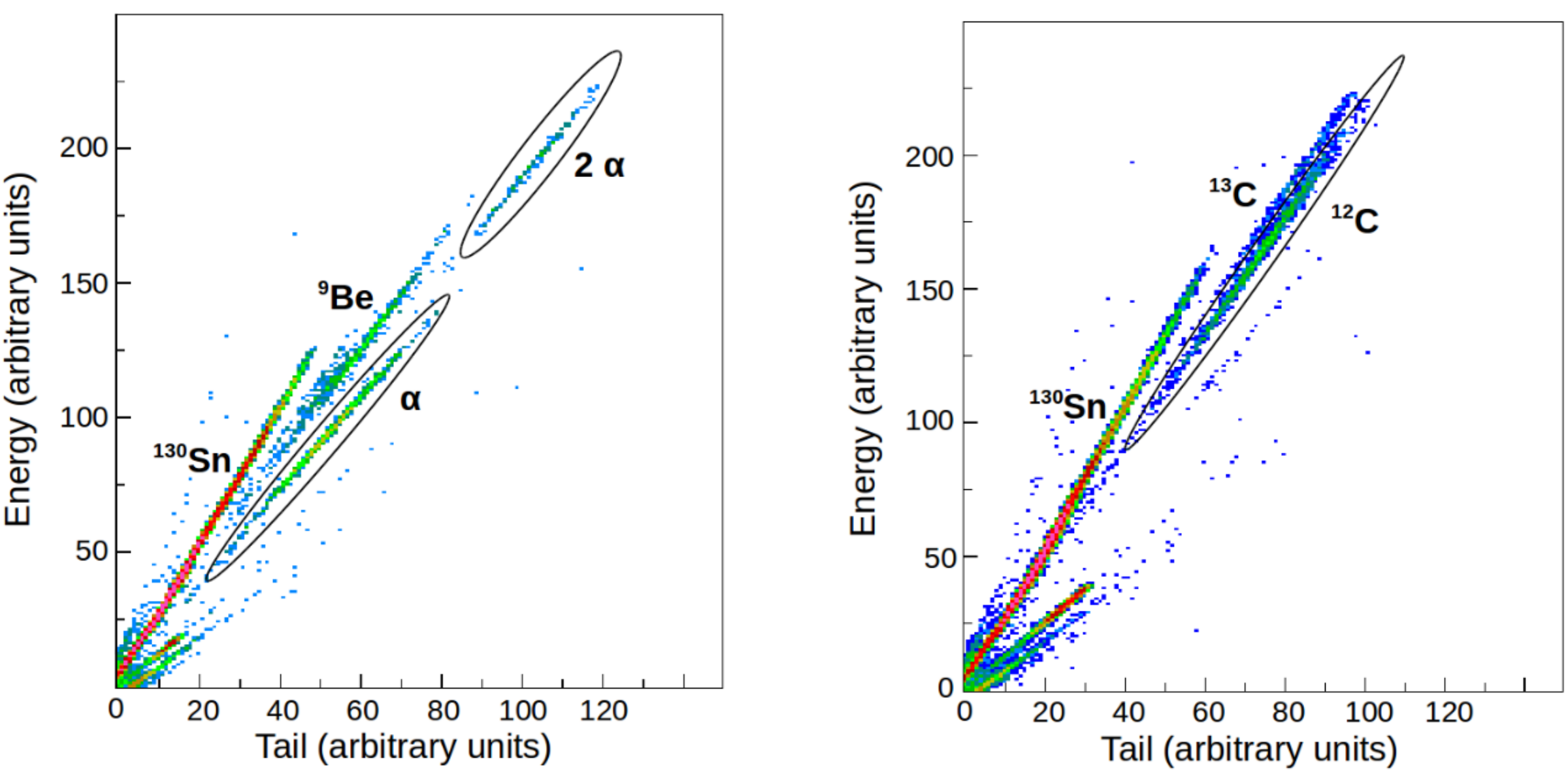}
\caption{Particle Identification spectra for reactions of $^{130}$Sn on a $^9$Be (left panel) and $^{13}$C (right panel) target.  The total energy deposited in a CsI crystal is plotted against the slow component (tail) of the signal to give separation between different light particles.  Unreacted $^{130}$Sn ions are sometimes scattered at large enough angles to be measured in BareBall.  The $\alpha$ and 2$\alpha$ groups are more cleanly separated in the $^9$Be target case than the $^{13}$C and $^{12}$C recoils in the $^{13}$C target case.}
\label{pid}
\end{figure}

In addition to the work using the (d,p) reaction on neutron-rich tin isotopes, our collaboration has also measured $\gamma$ rays emitted following the ($^9$Be,$^8$Be) and ($^{13}$C,$^{12}$C) reactions on short-lived beams of $^{130}$Sn and $^{132}$Sn \cite{All14}.  This method has the advantage of providing more precise excitation energies, access to the lifetimes of states, and information from the $\gamma$ decay cascades.  The emitted particles, 2$\alpha$ particles following the breakup of $^8$Be in the case of ($^9$Be,$^8$Be) and $^{12}$C for the ($^{13}$C,$^{12}$C) reaction, were detected in the BareBall CsI(Tl) array \cite{Gal10}.  Coincident $\gamma$ rays were measured in the CLARION array of 11 Compton suppressed, segmented, high-purity germanium clover detectors \cite{Gro00}. The Particle IDentification (PID) plot for the $^{130}$Sn beam experiment is shown in Figure \ref{pid}.  When both $\alpha$ particles are detected in the same CsI crystal, the combined signal is clearly separated in the PID plot, thus providing a clean tag for population of states in $^{131}$Sn.  Similarly, by requiring two single $\alpha$ signals, corresponding to events where the $\alpha$ particles are detected in neighboring crystals, a clean gate is available for analysis of the coincident $\gamma$ rays.  

These techniques are complimentary to using the (d,p) reaction, which also transfers a single neutron, but with different selectivity for populating single-particle like states.  In particular, the ($^{13}$C,$^{12}$C) reaction, compared to (d,p) and ($^9$Be,$^8$Be), preferentially populates states with $j=\ell + \frac{1}{2}$, as spin-flip transitions are preferred in nucleon transfer reactions, and the last neutron in $^{13}$C is in the $p_{1/2}$ orbital.  Hence, looking at relative intensities of individual states from these different reactions can aid in defining the $J^{\pi}$ of that state.

Placing silicon arrays inside large $\gamma$ ray arrays provides an opportunity to simultaneously gather information from the charged ejectile following a transfer reaction and the subsequent $\gamma$ rays.  The Oak Ridge Rutgers University Barrel Array (ORRUBA)\cite{Pai07} of position sensitive silicon detects can fit, in its entirety, inside the Gammasphere array of 110 high purity germanium detectors \cite{Lee90}.  The combined instrumentation, named the Gammasphere ORRUBA Dual Detectors for Experimental Structure Studies (GODDESS)\cite{Rat13}, has been used with beams provided by the ATLAS facility to measure reactions with $^{134}$Xe, $^{95}$Mo, and $^3$He beams\cite{Pai17}.  Work is in progress to combine ORRUBA with Gretina/GRETA \cite{Lee03} in a similar way to GODDESS, to make a powerful instrument for spectroscopic measurements using transfer reactions with reaccelerated beams at FRIB.

\subsection{Hybrid Array for Gamma Ray Detection (HAGRiD)}

The advantages of being able to measure $\gamma$ rays that follow transfer reactions are such that  our collaboration decided to build a scintillator array for this purpose, as well as for nuclear decay studies.  The array needed to be flexible, as it will be used in multiple configurations, and transportable, so that it can be used at various laboratories where these experiments are performed.  For these reasons scintillator detectors were preferred over germanium detectors, which require liquid nitrogen cooling and are prohibitively expensive.  Lanthanum bromide was chosen for its relatively good energy resolution ($<3\%$ at 662~keV), high intrinsic efficiency ($\sim35\%$ at 1~MeV) and good timing resolution ($\sim$ 200 ps).  The crystals selected are of the \textsc{B}ril\textsc{L}an\textsc{C}e 380$^{\textsc{TM}}$ type from Saint Gobain Crystals.  As of the time of writing, the HAGRiD array comprises of 27 2"x2" detectors, with a further 10 3"x3" crystals on order.  The crystals are currently being coupled to Hamamatsu R6231-100 photomultiplier tubes (PMTs) which include super bialkali photocathodes for increased quantum efficiency \cite{Nak10}.  

\begin{figure}
\includegraphics[width=4.in, trim={7cm 20cm 10cm 10cm},clip]{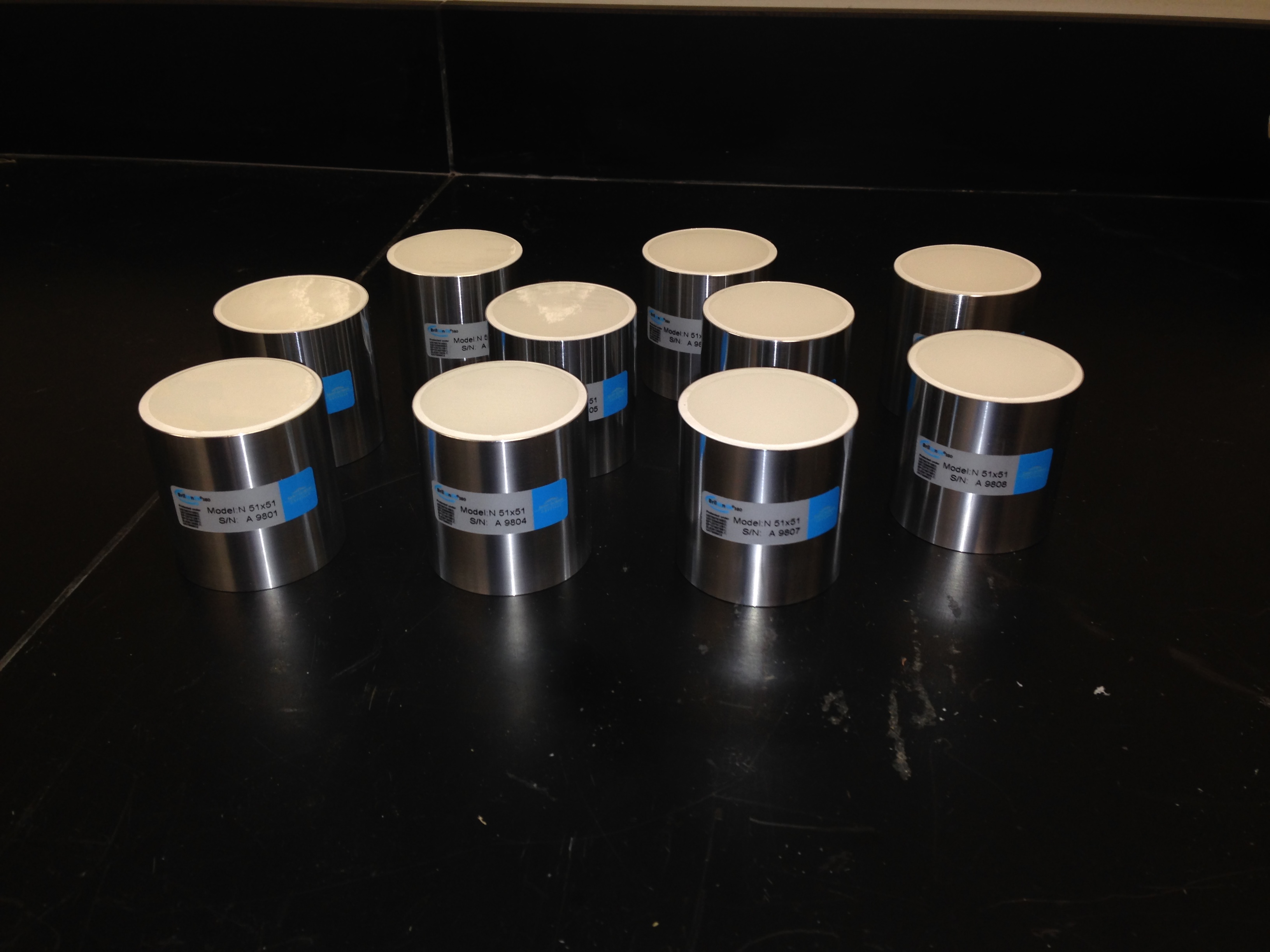}
\caption{The first 10 2" x 2" LaBr$_3$(Ce) crystals of the HAGRiD array were delivered in October 2015.}
\label{hagrid}
\end{figure}

The early implementation of HAGRiD, with 2"x2" crystals coupled to a combination of Hamamatsu R7724, R7224-100, and R6231-100 PMTs, has been used in experiments with the Jet Experiments in Nuclear Structure and Astrophysics (JENSA) gas jet target \cite{Bar15} at the National Superconducting Cyclotron Laboratory (NSCL) and in $\beta$-delayed neutron emission studies at the HRIBF.  A recent $\beta$-delayed neutron emission experiment at the NSCL with the Versatile Array of Neutron Detectors (VANDLE)\cite{Pau14, Pet16} used HAGRiD with the optimized R6231-100 PMT's in a temporary coupling. Once the array is completed, its flexibility to be positioned around the JENSA gas target, other $\gamma$ detectors, or neutron detectors will make it a valuable component of experiments studying the decay or reactions of exotic nuclei. 

\section{Proton Transfer}
The (d,n) reaction transfers a proton onto a nucleus in an analogous way to neutron transfer using the (d,p) reaction.  The difficulty in using the (d,n) reaction is with measuring neutrons in place of protons, which can be detected with relative ease and good resolution using silicon, or other charged-particle detectors.  Developing (d,n) reactions in inverse kinematics, such that they can be used with radioactive ion beams is important to placing single-proton spectroscopy on the same experimental footing as single-neutron spectroscopy.  

\begin{figure}
\includegraphics[width=4.in, trim={0cm 0cm 0cm 0cm},clip]{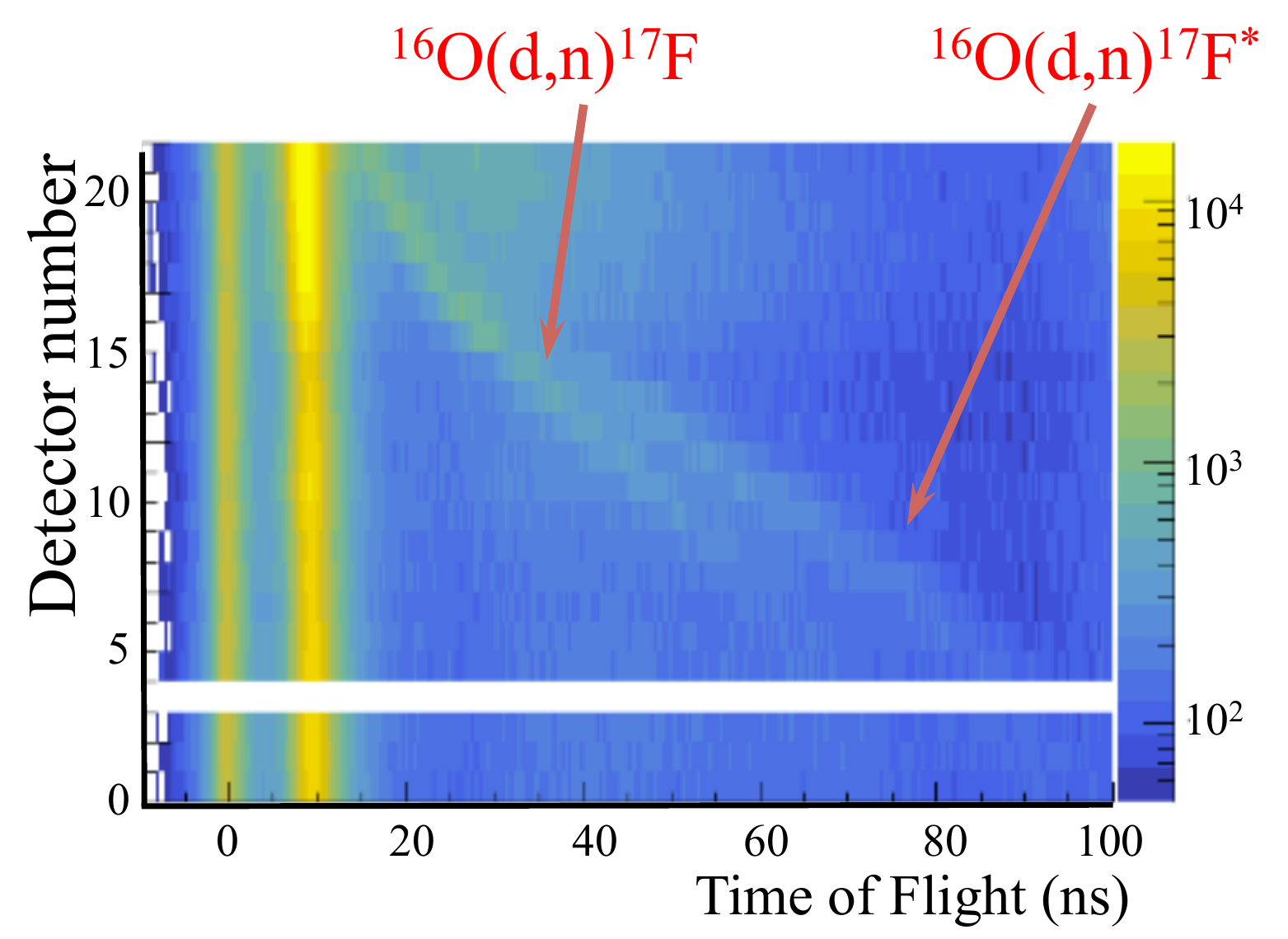}
\caption{Kinematics plot for the $^{16}$O(d,n)$^{17}$F reaction.  The detector number relates to the angle at which the neutron ejectile is measured, each detector covering roughly 3.4$^{\circ}$.}
\label{dn}
\end{figure}

We performed an experiment at the Nuclear Structure Laboratory (NSL), at the University of Notre Dame, using stable beams of $^{12}$C and $^{16}$O incident on a deuterated polyethylene target, with energies ranging from 23.5 to 41.7~MeV and at 64~MeV, respectively.  Neutrons emerging from the (d,n) reaction were detected in 21 bars of VANDLE, placed at 0.5~m from the target, and 5 bars containing a xylene-based liquid scintillator.  The time of flight of the neutrons in VANDLE was found from the difference between the timing signals from the bunched beam and VANDLE.   A scintillator detector, used as a beamstop, provided an additional timing signal.  The detectors were placed at angles from 60$^{\circ}$ (detector 21) to 170$^{\circ}$ (detector 1). The time of flight (ToF) is shown as a function of detector number, which varies smoothly with angle, in Fig. \ref{dn} for the $^{16}$O(d,n) measurement.  As VANDLE is made of plastic scintillator, a $\gamma$ ray background is present, including the prompt flash of $\gamma$'s when the beam hits the target, and a second, more intense flash from the beamstop.  The target-induced $\gamma$ flash gives a timing reference to calibrate the time of flight.

Two loci are apparent in Fig. \ref{dn}, the lower from neutrons emerging from the $^{16}$O(d,n) reaction populating the ground state of $^{17}$F, and the upper is from population of the 3.1~MeV state in $^{17}$F.  The resolution in excitation energy in $^{17}$F improves quadratically as the time of flight increases.  These data are still under analysis.   The next step in this program is to successfully measure the $^7$Be(d,n)$^8$B reaction using a $^7$Be beam from the TwinSol facility at the NSL.  The secondary beam will have a lower intensity than what was used for the stable beam test, so extracting events corresponding to the (d,n) reaction from the $\gamma$ ray background will be crucial.  We are working on methods of recoil identification to make these measurements possible.

\section{Summary}
Building on the work performed using transfer reactions with fragmentation, in-flight, and ISOL beams, our collaboration has developed new tools and techniques to extend their reach for the next generation of radioactive ion beam facilities, such as FRIB.  At this important moment, we have focused our efforts on developing techniques for measuring $\gamma$ rays in coincidence with emergent particles following transfer, measuring proton transfer using the (d,n) reaction, and using the results of transfer reactions to inform nuclear astrophysics simulations.  It is necessary that detector and target technology, experimental techniques, and theory continue to improve in parallel to radioactive beam developments in order to realize the full potential of new facilities.

\section*{Acknowledgments}

This material is based upon work supported by the U.S. Department of Energy, Office of Science, Office of Nuclear Physics under contract number DE-FG02-96ER40983 and DE-SC000174, and research using resources of the Holifield Radioactive Ion Beam Facility, which was a DOE Office of Science User Facility operated by Oak Ridge National Laboratory, which is managed by UT-Batelle, LLS, for the U.S. Department of Energy (DOE). This work was funded in part by the National Nuclear Security Agency, Stockpile Stewardship Academic Alliance, and by the National Science Foundation.  The author would like to thank the faculty, staff, and students at the NSL.

\bibliographystyle{ws-procs9x6} 
\bibliography{sanibel16-kate}

\end{document}